# LarQucut: A New Cutting and Mapping Approach for Large-sized Quantum Circuits in Distributed Quantum Computing (DQC) Environments

XINGLEI DOU, LEI LIU, ZHUOHAO WANG, PENGYU LI, Beihang University, China

## ABSTRACT

Distributed quantum computing (DQC) is a promising way to achieve large-scale quantum computing. However, mapping large-sized quantum circuits in DQC is a challenging job; for example, it is difficult to find an ideal cutting and mapping solution when many qubits, complicated qubit operations, and diverse QPUs are involved. In this study, we propose LarQucut, a new quantum circuit cutting and mapping approach for large-sized circuits in DQC. LarQucut has several new designs. (1) LarQucut can have cutting solutions that use fewer cuts, and it does not cut a circuit into independent sub-circuits, therefore reducing the overall cutting and computing overheads. (2) LarQucut finds isomorphic sub-circuits and reuses their execution results. So, LarQucut can reduce the number of sub-circuits that need to be executed to reconstruct the large circuit's output, reducing the time spent on sampling the sub-circuits. (3) We design an adaptive quantum circuit mapping approach, which identifies qubit interaction patterns and accordingly enables the best-fit mapping policy in DQC. The experimental results show that, for large circuits with hundreds to thousands of qubits in DQC, LarQucut can provide a better cutting and mapping solution with lower overall overheads and achieves results closer to the ground truth.

## 1 INTRODUCTION

Quantum computing is a revolutionary approach that can provide significant speedups over classical computers for many problems [1,2,3,18,35,40]. Yet, to solve a real problem in practice, a quantum computer with at least thousands of qubits is expected [4,7]. For example, it is estimated that 4000 logical qubits are required to crack RSA-2048 [4]. Most of today's quantum computers have only tens to hundreds of available qubits [5,6]. These devices are NISQ computers [7], which are easily disturbed and prone to errors. It is difficult to scale up the number of qubits on a specific quantum processing unit (QPU) due to fabrication challenges [8,9]. Distributed quantum computing (DQC) is a practical approach that can scale up quantum computers in the NISQ era [9,10,12,16,21,33,34]. In DQC, many small-sized QPUs work coordinately via inter-node communications to form a large quantum computing system. Inter-node communications between QPUs (remote gates) rely on remote EPR (Einstein-Podolsky-Rosen) pairs generated by communication qubits [11]. Inter-node communications are much more error-prone and time-consuming than in-node communications between data qubits within a QPU [9,10,16]. Therefore, the number of inter-node communications should be carefully optimized by leveraging quantum software, e.g., OS, compiler, etc.

Quantum circuit cutting is an approach to decompose a large-sized quantum circuit into smaller sub-circuits. Then, the sub-circuits can be mapped on different QPUs separately, and the result of the original large circuit can be obtained by combining the sub-circuits' execution results [8,12]. Moreover, circuit cutting can be used to reduce inter-node communications in DQC because it removes the interactions between qubit groups. A quantum circuit can be decomposed using wire cuts and gate cuts [12,13]. A wire cut conducts on a specific qubit wire. It introduces a new qubit wire and moves the remaining operations after the cut to the new wire [14]. A gate cut conducts on a specific two-qubit gate, i.e., the two-qubit gate is computed as two single-qubit gates in separate sub-circuits [15]. More details on wire cuts and gate cuts can be found in Section 2.2. The classical post-processing overheads grow exponentially with the number of wire/gate cuts [12,13]. Thus, the number of cuts should be minimized in practice.

Authors' Contact Information: X. Dou, L. Liu (Corresponding author), Z. Wang, P. Li, Beihang University, No. 37 Xueyuan Road Zhongguancun, Haidian District, Beijing, China, 100191, China; e-mails: lei.liu@zoho.com, liulei2010@buaa.edu.cn.

Cutting and mapping a large-sized quantum circuit can be quite different compared with handling small-sized circuits. A straightforward way is to cut the large-sized circuit into completely independent sub-circuits that can run on small-sized QPUs [8,12,13,22,38,39]. Doing so can effectively avoid inter-node communications between QPUs, but it would need a large number of wire/gate cuts, especially for circuits that have dense pair-wise interactions between qubits (e.g., Quantum Fourier Transform, QFT). And, doing in this way incurs high classical post-processing overheads when reconstructing the original circuit's result. By contrast, This paper has the following new insights on this topic. (1) We find that it is not necessary to cut a circuit into completely independent sub-circuits. More cuts lead to higher classical post-processing overheads. To reduce this overhead and have the ideal solutions, this work does not always cut a circuit into completely independent sub-circuits. Instead, it only cuts the wires/gates that can effectively reduce the interactions between qubit groups. Compared with prior studies [8,13,22] that cut a circuit into independent sub-circuits, we can use fewer wire/gate cuts and have lower classical post-processing overheads. (2) A better cutting solution often needs to use wire cuts and gate cuts cooperatively. A wire cut removes the dependency between gates conducting on a specific logical qubit. It is effective in decomposing circuits like Bernstein-Vazirani (BV), where many two-qubit gates share one logical qubit. A gate cut removes the interaction between two logical qubits and does not introduce new qubits. It is used to cut remote gates between qubits mapped on separate QPUs. For a large-sized circuit that has complicated interactions between qubits, wire cuts and gate cuts should be used cooperatively to effectively decompose the circuit. We find no single approach can handle diverse cases. (3) Regarding qubit mapping across QPUs, no matter how to cut the large circuits, there are often cases where the number of qubits in a specific sub-circuit exceeds the number of qubits that a specific QPU can provide. Thus, qubits with frequent interactions should be mapped or swapped onto the same QPU. Cross-node mapping often leads to inter-node communications between qubits. In many cases, we avoid mapping a group of qubits with frequent interactions across QPUs, and when a logical qubit has few interactions with local qubits but many interactions with remote qubits on the other node, it should be moved to the other node using remote SWAPs.

To this end, we propose LarQucut, a new quantum circuit cutting and mapping approach in DQC environments. LarQucut has several new designs. (1) It removes the interactions between qubit groups by employing wire cuts and gate cuts cooperatively. It has low classical post-processing overheads because it does not have to cut a circuit into completely independent segments (i.e., hemicut). (2) The execution results of isomorphic sub-circuits can be reused when reconstructing the original circuit's result [17]. Therefore, LarQucut cuts the circuit into isomorphic sub-circuits and reuses their execution results to reduce the classical post-processing overheads. (3) LarQucut has an adaptive quantum circuit mapping approach, including two initial mapping policies (i.e., hotness-mapping and weakness-mapping) that work well for diverse circuits. It also has a new heuristic function that moves qubits with frequent interactions to the same DQC node using remote SWAPs. We compare LarQucut with the most related studies [8,24,36] using circuits that have up to 1024 qubits. LarQucut can find cutting solutions that use 22.3%~32.1% fewer cuts than existing circuit-cutting approaches [8,24]. When mapping circuits across QPUs, LarQucut reduces the number of consumed EPR pairs by 38.1%. Furthermore, LarQucut reduces up to 27.7% in terms of the absolute error compared with the ground truth, indicating that LarQucut brings a higher fidelity. To sum up, this paper makes the following contributions.

**(1) A key observation.** We observe that not all cuts are equally effective for having the ideal circuit cutting and mapping solutions in DQC. For a large-sized quantum circuit, there often exist several wires/gates that are critical. Only cutting these critical wires/gates can effectively reduce the number of EPR pairs for inter-node communications, improve the fidelity and bring a low overhead, outperforming the existing approaches that often involve more cuts that might bring higher overheads.

**(2) Hemicut that does not cut circuits into completely independent sub-circuits.** Cutting a circuit into completely independent sub-circuits requires a large number of cuts, especially for circuits with dense interactions between qubits. A circuit does not need to be further cut when there are only a few inter-node communications left that can be further optimized during circuit mapping (e.g., the remaining remote gates that use the same logical qubit can be executed using one EPR pair). This approach reduces the classical post-processing overheads by using fewer cuts.

**(3) Cutting and mapping framework with isomorphic sub-circuits reusing.** We design a quantum circuit-cutting

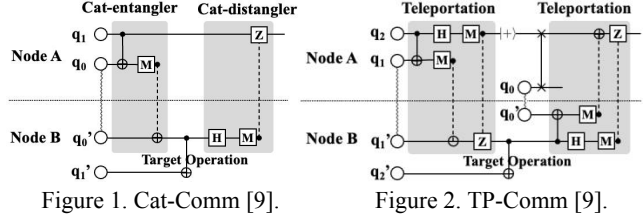

Figure 1. Cat-Comm [9].    Figure 2. TP-Comm [9].

approach that quickly searches critical wire cuts and gate cuts to provide ideal solutions. We further design an isomorphic sub-circuits reusing mechanism that cuts a circuit into isomorphic sub-circuits when possible and reuses their execution results in classical post-processing. For mapping quantum circuits across QPUs, we design adaptive initial mapping policies that work well for diverse circuits and a new heuristic function that reduces the EPR pairs required for inter-node communications by moving qubits with frequent interactions to the same DQC node.

**(4) Real implementation and evaluations across large-sized quantum circuits.** We implement LarQucut based on Qiskit 1.0.2 [23] and Circuit Knitting Toolbox 0.7.0 [24]. The experiments are conducted on noisy superconducting QPUs simulated using Qiskit-aer [23] and Diskit [25] based on real noise data from IBM. We compare LarQucut with state-of-the-art approaches using large-sized quantum circuits with up to 1024 qubits and show the advantages.

## 2  BACKGROUND AND MOTIVATION

In this section, we introduce critical background knowledge of DQC and quantum circuit cutting. We further show the key challenges that need to be addressed.

### 2.1  Distributed Quantum Computing

Quantum data cannot be replicated due to the no-cloning theorem [11]. So, the inter-node communications between two QPUs rely on the remote EPR entanglement in a pair of qubits. A remote EPR pair contains two entangled qubits with the state of $(|00\rangle + |11\rangle) / \sqrt{2}$. It is generated using communication qubits on two QPUs. Communication qubits are specifically designed to generate EPR pairs. Other qubits on a DQC node are data qubits, which store the logical qubits. There are two common protocols for inter-node communications, i.e., Cat-Comm and TP-Comm. Figure 1 and Figure 2 show how they perform. More details on Cat-Comm and TP-Comm can be found in [9,10,16]. Consecutive remote gates sharing the same logical qubit can be executed using one EPR pair [9]. However, involving more EPR pairs increases communication overheads and also impairs fidelity [9,10,16].

### 2.2  Quantum Circuit Cutting

**Wire cut.** When each QPU has three qubits, the 4-qubit GHZ state circuit in Figure 3 needs to be mapped across QPUs, incurring at least one remote gate. Figure 3 shows how to cut this circuit into two sub-circuits using a wire cut. To obtain the result of the original circuit, we need to run sub-circuit 1 with $q_1$'s observable measured in four Pauli matrix bases (i.e., I, X, Y, Z), and run sub-circuit 2 with the state of $q_1$' initialized to four eigen states (i.e., $|0\rangle, |1\rangle, |+\rangle, |i\rangle$). Finally, we reconstruct the output of the original circuit $\rho$ according to $\rho = (A_1+A_2+A_3+A_4) / 2$ [14], where

$$A_1 = \mathrm{Tr}(\rho I)[|0\rangle\langle 0| + |1\rangle\langle 1|]$$
$$A_2 = \mathrm{Tr}(\rho X)[|0\rangle\langle 0| - |1\rangle\langle 1|]$$
$$A_3 = \mathrm{Tr}(\rho Y)[2|+\rangle\langle +| - |0\rangle\langle 0| - |1\rangle\langle 1|]$$
$$A_4 = \mathrm{Tr}(\rho Z)[2|i\rangle\langle i| - |0\rangle\langle 0| - |1\rangle\langle 1|]$$

Each trace operator Tr () represents running sub-circuit 1 and measuring the output in one of the four Pauli bases. Each density matrix $|x\rangle\langle x|$ indicates running sub-circuit 2 with the state initialized in one of the four eigen states. Four pairs of Kronecker products are performed according to the above equation. Therefore, the classical post-processing overhead of wire cuts is $O(4^k)$, where k is the number of wire cuts. Notably, since measuring a qubit in I and Z bases results in the same circuit, three variants of sub-circuit 1 and four variants of sub-circuit 2 are generated.

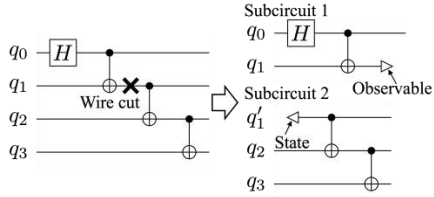 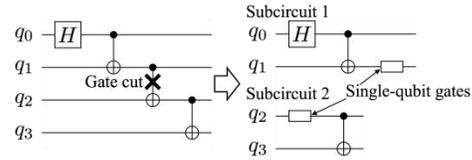

Figure 3. Cutting a GHZ state circuit using a wire cut.    Figure 4. Cutting a GHZ state circuit using a gate cut.

**Gate cut.** Figure 4 shows how to cut the GHZ circuit using a gate cut. According to [15], a quantum gate in the form $e^{i\theta A_1 \otimes A_2}$ with $A_1^2=I$ and $A_2^2=I$ can be decomposed into six single-qubit operations, as shown in the following equation.

$$S(e^{i\theta A_1 \otimes A_2}) = \cos^2\theta S(I \otimes I) + \sin^2\theta S(A_1 \otimes A_2) \\ + \frac{1}{8}\cos\theta\sin\theta \sum_{\alpha_1,\alpha_2 \in \{\pm1\}^2} \alpha_1\alpha_2[S((I+\alpha_1 A_1) \otimes (I+i\alpha_2 A_2)) \\ + S((I+i\alpha_1 A_1) \otimes (I+\alpha_2 A_2))]$$

S (U) denotes a super operator that can transform the quantum state ρ to $U\rho U^\dagger$ by applying the gate U, i.e., S (U) ρ = $U\rho U^\dagger$. To obtain the result of the original circuit, we need to replace the CNOT gate that has been cut with a single-qubit gate for each sub-circuit. Six pairs of Kronecker products are performed according to the above equation. The classical post-processing overhead of gate cut is O ($6^k$), where k is the number of gate cuts. We recommend [15] for more details on gate cutting.

**Quantum circuit cutting in DQC.** Quantum circuit cutting in DQC makes large-sized quantum computing feasible. For monolithic quantum chips that do not support inter-node communications, it is expensive to cut a large-sized quantum circuit and map the sub-circuits. This is because the large-sized quantum circuit has to be cut into completely independent sub-circuits to be mapped on these devices, which often requires a large number of cuts and incurs huge computational overheads. In the context of DQC, inter-node communications can happen between qubits mapped across two QPUs via remote EPR pairs. Thus, the large-sized quantum circuit does not have to be cut into completely independent sub-circuits. This reduces the number of cuts in circuit cutting, making it easier to evaluate larger-sized quantum circuits. In this work, the maximum number of qubits that LarQucut can evaluate is limited, primarily due to the exponential computational overheads to simulate the sub-circuits on classical computers. As detailed in Section 5.3, simulation takes over 99% of the total execution time. However, when practical DQC hardware becomes available in the near future, sub-circuit sampling can have results in a much shorter time, and LarQucut will be able to make the execution of quantum circuits with a larger number of qubits effective.

## 2.3  Key Challenges

How to design an ideal scheme for cutting and mapping a large-sized circuit in DQC? The latest studies optimize the circuit mapping algorithms in DQC [26,27,28] and reduce inter-node communications by increasing the information transferred per EPR pair [9,10,16]. Some studies cut a large circuit into smaller sub-circuits that fit small QPUs in DQC [8,12,22]. The work in [12] reuses qubits leveraging mid-circuit measurement and can effectively reduce the qubit usage in circuit cutting. In terms of mapping/cutting a large circuit, the problem might be more challenging because large-sized circuits have complicated interactions between qubits. We summarize the challenges of circuit cutting and mapping for large-sized circuits in DQC as follows.

(1) A large-sized quantum circuit has a large number of qubits with complicated interactions. When decomposing a large-sized circuit using circuit cutting, there are many wire cut and gate cut options available. These cuts are not equally effective in decomposing the circuit. There are many available cutting solutions that cut on different wires/gates in the circuit. These solutions differ in the sub-circuit sizes, number of inter-node communications, and classical post-processing overheads. Therefore, the search space for circuit cutting is large and complicated. Besides, QPUs in a DQC system may differ in the number of qubits and qubit topology, making it challenging to find an ideal mapping solution that all sub-circuits can be well matched with the QPUs.

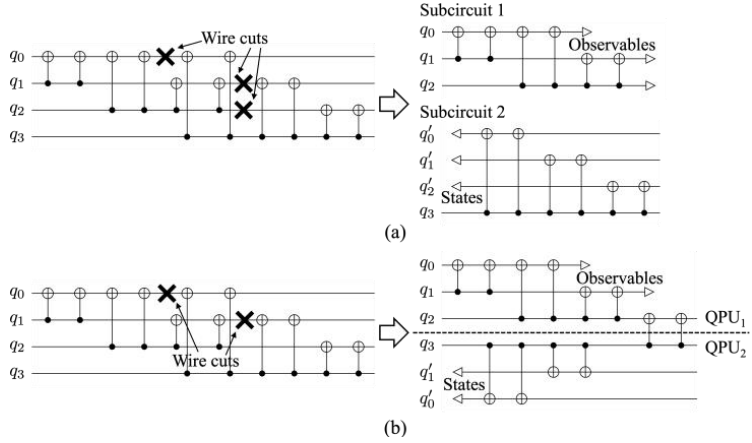

Figure 5. An example of cutting 4-qubit QFT circuit to map it on 3-qubit QPUs. Single-qubits are not shown here as they do not incur inter-node communications. If the circuit is not cut, it needs to execute at least six remote gates. (a) A cutting solution that cuts the circuit into two independent sub-circuits using three wire cuts. (b) A better cutting solution that only cuts two critical wires.

(2) In terms of circuit cutting, there are many available cutting solutions for a specific large-sized quantum circuit. Some solutions use many cuts and therefore incur high computational overheads. Some use fewer cuts, but the number of qubits in the sub-circuits exceeds that of the QPU, i.e., the sub-circuits have to be mapped across QPUs and still incur many inter-node communications. Moreover, in some studies, the circuit is cut into completely independent sub-circuits. Among all cuts used in these solutions, some cuts are unnecessary because they can remove only a few inter-node communications but incur higher computational overheads. Figure 5 illustrates an example of cutting a 4-qubit QFT circuit to map it on 3-qubit QPUs. Figure 5-(a) shows a cutting solution that cuts the circuit into two independent sub-circuits using three wire cuts. The number of qubits in sub-circuit 2 exceeds that of the 3-qubit QPU. So, sub-circuit 2 has to be mapped across QPUs, and at least two remote gates are required. Figure 5-(b) illustrates a better solution that only cuts two critical wires. It also requires two remote gates but has lower classical post-processing overheads because it uses fewer cuts. The wire cut on $q_2$ in Figure 5-(a) is unnecessary because it does not reduce the inter-node communications but incurs higher computational overheads.

(3) Regarding circuit mapping in DQC, the key challenge is to minimize the number of EPR pairs required for inter-node communications. There are often cases where a sub-circuit is larger than the QPU and has to be mapped across QPUs. If qubits with frequent interactions are not mapped on the same QPU, they incur many remote operations that are error-prone and time-consuming, which degrades fidelity. The interaction pattern between qubits of a circuit may change after a number of gates are executed. Therefore, there might be cases where a qubit has few interactions with local qubits but many interactions with remote qubits on the other node, incurring many error-prone inter-node communications. Moreover, EPR pairs in DQC are not effectively utilized. Even though there are many remote gates sharing the same logical qubit in a circuit, they are often not consecutive and cannot be executed sharing the same EPR pair.

## 3 CUTTING LARGE-SIZED QUANTUM CIRCUITS

To this end, we propose LarQucut, a new quantum circuit cutting and mapping mechanism for large-sized quantum circuits in DQC environments. For a specific large-sized quantum circuit, LarQucut first removes interactions between groups of qubits by cutting critical wires/gates. Then, it enables appropriate mapping policies to map the sub-circuits across QPUs (detailed in Section 4). The major challenge in cutting a large-sized quantum circuit is to reduce the circuit-cutting overheads. This section shows how LarQucut addresses this challenge by cutting critical wires/gates and leveraging isomorphism in large-sized quantum circuits.

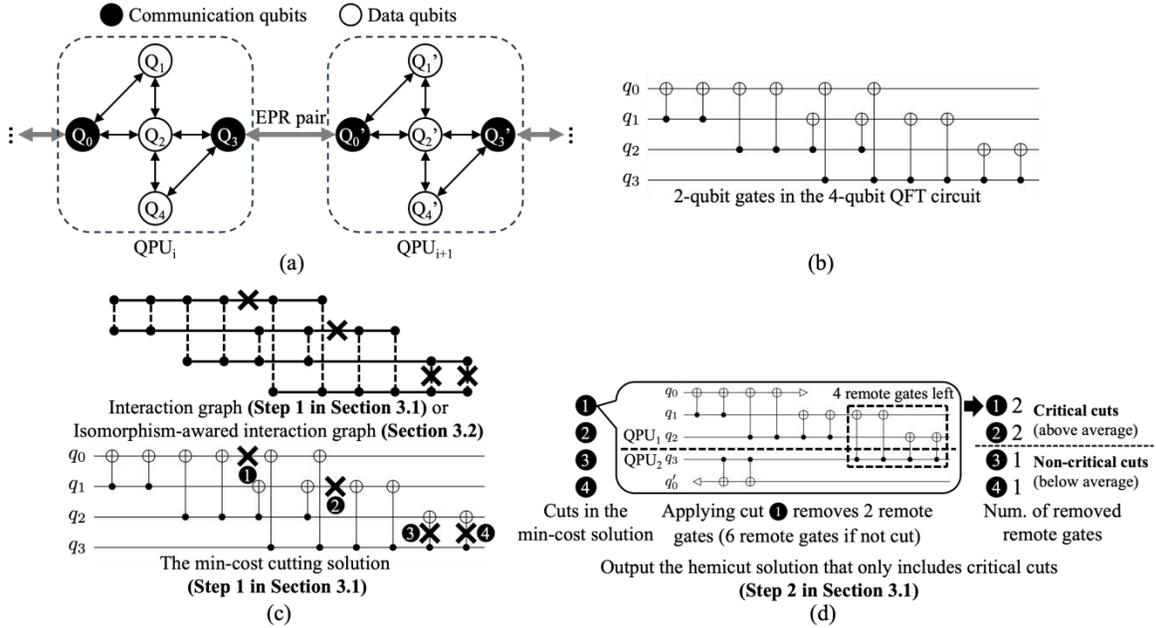

Figure 6. An example of LQ-hemicut's workflow. (a) The DQC model. (b) The 4-qubit QFT circuit. Single-qubit gates are not shown here as they do not incur inter-node communications. (c) LQ-hemicut constructs an interaction graph and searches the min-cost cutting solution using the interaction graph. When LQ-reuse (Section 3.2) is enabled, the isomorphism-aware interaction graph is used instead. (d) LQ-hemicut identifies critical cuts and outputs the hemicut solution that only includes the critical cuts.

The first key design is a topology-aware quantum circuit-cutting approach, i.e., LQ-hemicut in Section 3.1. "Hemicut" indicates that this approach does not have to cut a circuit into completely independent sub-circuits. It saves the cuts required to cut a large-sized circuit. Besides, LQ-hemicut leverages the DQC topology information to search for critical cuts in the large search space. Using the DQC topology information, LQ-hemicut can have the number of inter-node communications that need to be executed when a circuit is mapped across QPUs. It identifies critical cuts that can effectively reduce inter-node communications and only uses them in the final cutting solution.

The second key design is to reuse the execution results of isomorphic sub-circuits during the classical post-processing, i.e., LQ-reuse in Section 3.2. When a large-sized quantum circuit is cut into isomorphic sub-circuits, the execution results of these isomorphic sub-circuits can be reused, reducing the number of sub-circuits that need to be executed (i.e., the sampling overhead). LQ-reuse identifies isomorphic sub-circuits in a large-sized quantum circuit and cuts the circuit into isomorphic sub-circuits when possible.

### 3.1 Cutting Large-sized Circuits using LQ-hemicut

Figure 6 shows an example that illustrates how LQ-hemicut cuts a large-sized quantum circuit. Figure 6-(a) shows the DQC model, which has several 5-qubit IBM Yorktown QPUs arranged in the 1D nearest-neighbor topology. For each QPU, two physical qubits are reserved as the communication qubits (highlighted in black). Logical qubits can only be mapped on the remaining data qubits. Figure 6-(b) shows the 4-qubit QFT circuit. Figure 6-(c) shows that LQ-hemicut constructs an interaction graph (or an isomorphism-aware interaction graph detailed in Section 3.2) for the QFT circuit. Then, it finds a cutting solution that cuts the circuit into independent sub-circuits using the graph. The details are in Step

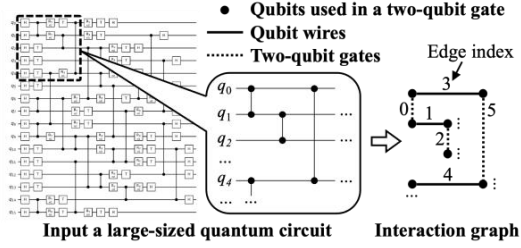

Figure 7. Interaction graph of a large-sized quantum circuit.

1. Figure 6-(d) shows that LQ-hemicut identifies critical cuts and only applies critical cuts in the final solution. (Details are in Step 2). The detailed workflow of LQ-hemicut is shown below.

**Step 1 - Searching for a cutting solution for the large-sized quantum circuit according to the DQC topology.** This process includes three sub-steps. (1) LQ-hemicut converts the circuit cutting problem to an equivalent graph partitioning problem by constructing an interaction graph for the input circuit. Figure 7 shows an example where LQ-hemicut constructs an interaction graph for a large-sized circuit. A node in the graph represents a logical qubit used in a two-qubit gate. The solid horizontal edges represent qubit wires; the dashed vertical edges represent two-qubit gates. Cutting an edge in the interaction graph represents cutting a wire/gate in the circuit. In the example of Figure 6, the interaction graph of the 4-qubit QFT circuit is shown in Figure 6-(c).

(2) LQ-hemicut uses a boolean list s to represent a cutting solution candidate that cuts the interaction graph, where $s_i$ = true means the i-th edge in the interaction graph is cut, $s_i$ = false means the i-th edge is not cut. The length of s represents the number of edges for which cutting decisions have been made. LQ-hemicut initializes a min-heap that stores the cutting solution candidates. The min-heap stores the candidate with the lowest cost in the root node. A lower cost value represents that the candidate uses fewer cuts and incurs fewer inter-node communications. LQ-hemicut adds an empty list (i.e., no cutting decisions have been made) to the min-heap as the initial solution.

(3) LQ-hemicut searches for a cutting solution from the candidates. It iterates to update the min-cost solution until the min-heap of cutting solution candidates is empty. In each iteration, a cutting solution candidate is taken from the min-heap. If the candidate can not be used, LQ-hemicut skips it and proceeds to the next iteration. By doing this, LQ-hemicut avoids spending time on candidates that do not have better results, reducing the search space, especially when cutting a large-sized quantum circuit. Three types of cutting solution candidates can not be used. (i) If one of the sub-circuits has more qubits than the original circuit when a cutting solution candidate is applied, the candidate can not be used. (ii) If the cost of a candidate exceeds that of the min-cost solution, the candidate can not be used. This is because further iterations may lead to even higher costs and will not generate a better solution than the min-cost solution. (iii) If the cutting decisions for all edges are made, and the circuit is not cut into independent sub-circuits, the candidate can not be used.

When the cutting solution taken from candidates is a goal solution (i.e., the cutting decisions for all edges have been made, and the circuit is cut into independent sub-circuits), we update the min-cost solution if the new solution has a lower cost value. The cost is represented as a tuple that has four items. The first item is the minimum number of remote gates required when the sub-circuits are mapped across QPUs. With this item, the cutting solutions that cut more critical gates/wires can have a lower (better) cost value. LQ-hemicut gets this item by generating initial mappings for these sub-circuits using the initial mapping policies in Section 4.2. The second item is the sampling overhead (i.e., how many times the sub-circuits need to be executed). When there are isomorphism sub-circuits, the sampling overhead is reduced because the execution results of isomorphic sub-circuits can be reused. More details on reusing isomorphic sub-circuits are in Section 3.2. The third item is the classical post-processing overhead. It is calculated as $4^{k1}6^{k2}$, where k1 denotes the number of wire cuts, k2 denotes the number of gate cuts. With the second and the third items, LQ-hemicut can find cutting solutions that use fewer cuts and incur lower overheads. The last item is -1 × the search depth (i.e., the length of the cutting solution s). This item encourages LQ-hemicut to explore the candidates that have made cutting decisions for more edges. The min-heap sorts the candidates by comparing the values in the tuple. If low-index values are equal, the higher-index values are compared.

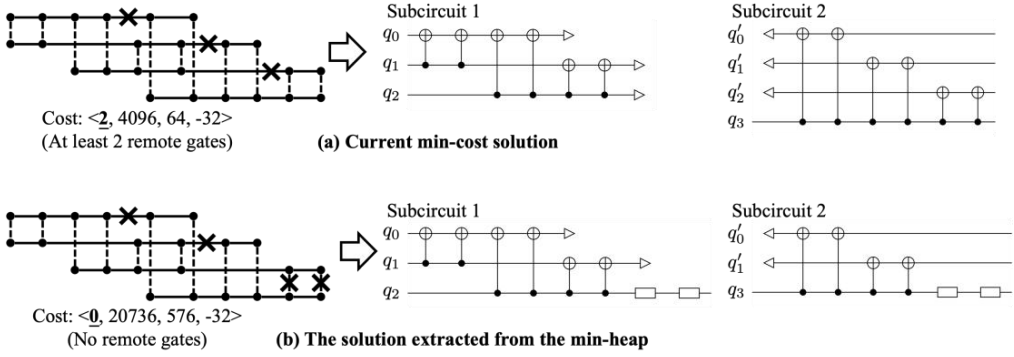

Figure 8. (a) The current min-cost solution for the 4-qubit QFT circuit in Figure 6. (b) The solution taken from the min-heap incurs less inter-node communications. It is used as the new min-cost solution.

For a solution that has not made cutting decisions for all edges (i.e., the length of s is less than the number of edges), LQ-hemicut generates two new candidates by adding a True and a False to the solution, respectively, indicating whether the next edge is cut or not. The two new candidates are put in the min-heap. When the iterations stop, we have the cutting solution candidate with the minimum cost value. This solution cuts the circuit into independent sub-circuits.

An example in Figure 8 shows the above process. Figure 8-(a) shows the current min-cost cutting solution when the 4-qubit QFT circuit is executed on QPUs with 3 data qubits in Figure 6. The min-cost cutting solution cuts the circuit into a 3-qubit sub-circuit 1 and a 4-qubit sub-circuit 2. Mapping sub-circuit 2 on the DQC system in Figure 6-(a) requires at least two remote gates. By contrast, a new solution from the min-heap cuts the circuit into two 3-qubit sub-circuits, which do not incur any remote gates (Figure 8-(b)). The new solution has a lower value of the first item in the cost tuple. Thus, the new solution is used as the min-cost solution. Some cuts in the solution are unnecessary because they only remove a few inter-node communications but incur high computational overheads. We remove these non-critical cuts from the cutting solution in the next step.

**Step 2 - Removing non-critical cuts from the final cutting solution.** In this step, LQ-hemicut identifies critical cuts in the min-cost solution and outputs the final cutting solution that only includes these critical cuts, i.e., the final cutting solution does not have to cut a circuit into completely independent sub-circuits. To achieve this, LQ-hemicut calculates the number of remote gates that can be removed by using each cut. Figure 6-(c) shows the min-cost cutting solution for the 4-qubit QFT circuit mapped on QPUs with 3 data qubits. When the circuit is not cut, it incurs at least six remote gates. The min-cost solution cuts the circuit into two independent sub-circuits using two wire cuts and two gate cuts. Figure 6-(d) shows the circuit when only cut ❶ is used. It requires at least four remote gates to map the circuit across QPUs. Thus, the number of remote gates that can be removed by using cut ❶ is calculated as 6 - 4 = 2. Similarly, when only one cut is used, the number of remote gates that can be removed is 2 for cut ❷, 1 for cut ❸, and 1 for cut ❹. We calculate the average number of remote gates that can be removed using these cuts. When the number of remote gates that can be removed by using a cut is larger than or equal to the average value, the cut is critical. Otherwise, the cut is unnecessary and should not be included in the final cutting solution. LQ-hemicut outputs the final cutting solution that only includes the critical cuts.

### 3.2 Reusing Isomorphic Sub-circuits in LQ-reuse

This section shows how LQ-reuse identifies isomorphic sub-circuits in a large-sized circuit and reuses them. LQ-reuse constructs an isomorphism-aware interaction graph that facilitates the LQ-hemicut to cut a large-sized quantum circuit into isomorphic sub-circuits. If LQ-reuse is enabled, the identification of isomorphic sub-circuits happens before circuit cutting, and the isomorphic-aware interaction graph is used in Figure 6-(c). LQ-reuse has two steps.

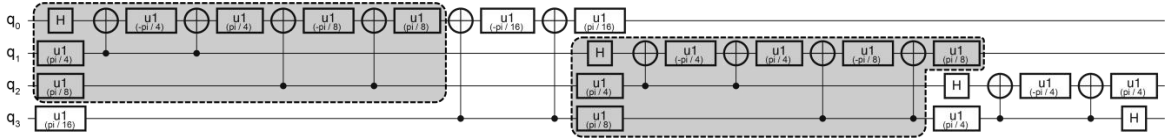
Figure 9. Isomorphic sub-circuits in 4-qubit QFT circuit are highlighted in gray.

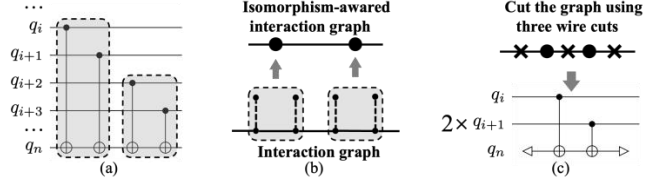
Figure 10. (a) Isomorphic sub-circuits in a large-sized BV circuit. (b) Construct isomorphism-aware interaction graph by contracting nodes of a isomorphic sub-circuit to a super node. (c) The two isomorphic sub-circuits generated by cutting the isomorphism-aware interaction graph using three wire cuts.

(1) Identifying isomorphic sub-circuits. LQ-reuse can identify a group of isomorphic sub-circuits that have no intersecting quantum gates from a large-sized quantum circuit. To avoid cross-node mappings, LQ-reuse only searches for isomorphic sub-circuits with a qubit count that does not exceed the largest QPU. LQ-reuse employs the VF2++ algorithm [29] for sub-circuit matching. It finds the isomorphic sub-circuits in a greedy way. LQ-reuse randomly selects a node from the interaction graph and uses it as the identified sub-circuit that has isomorphisms. Then, LQ-reuse expands this sub-circuit iteratively by selecting a neighboring node from the interaction graph. A node is added to the sub-circuit only if the expanded sub-circuit still has isomorphisms. When the sub-circuit can not be further expanded, LQ-reuse employs the VF2++ algorithm to obtain the matching solution that includes the sub-circuit and all of its isomorphisms. This process is repeated for N (default as 10) times, and LQ-reuse has N matching solutions. LQ-reuse selects the one that minimizes the number of cuts required to cut the large-sized circuit into isomorphic sub-circuits, i.e., the number of edges connecting to nodes of the isomorphic sub-circuits is the minimum. LQ-reuse can identify isomorphic sub-circuits in many quantum circuits. For example, Figure 9 highlights the isomorphic sub-circuits found in the 4-qubit QFT circuit in gray. LQ-reuse is not applicable for circuits that have multiple rotation blocks with arbitrary rotation angles (e.g., hardware-efficient ansatz), but it is especially effective for circuits with repetitive interaction patterns (e.g., Bernstein Vazirani algorithm, Greenberger-Horne-Zeilinger state, and Linear-cluster state, etc.). For example, Figure 10 shows that LQ-reuse can find isomorphic sub-circuits in a large-sized BV circuit.

(2) Contracting nodes of isomorphic sub-circuits. For each isomorphic sub-circuit, LQ-reuse contracts all nodes in the interaction graph corresponding to the isomorphic sub-circuit to a super node, as shown in Figure 10-(b). In this process, LQ-reuse replaces all nodes corresponding to an isomorphic sub-circuit with a super node. All edges that were connected to the removed nodes are now connected to the super node. The super node can not be cut into smaller sub-graphs, indicating that the wires and gates in the isomorphic sub-circuits will not be cut during the circuit cutting. This makes it easier for LQ-hemicut to cut the original circuit into isomorphic sub-circuits. For a large-sized quantum circuit, contracting nodes of isomorphic sub-circuits reduces the number of edges in the interaction graph, reducing the search space and speeding up the search process. In Figure 10-(c), two isomorphic sub-circuits are generated by cutting the updated interaction graph using three wire cuts. Only one of them needs to be executed when reconstructing the original circuit's result, i.e., the number of sub-circuits that need to be executed decreases. For an isomorphic sub-circuit shown in Figure 10-(c), LQ-reuse initializes the state in one of four eigen states (i.e., $|0\rangle$, $|1\rangle$, $|+\rangle$, $|i\rangle$) and measure the output in one of four Pauli bases (i.e., I, X, Y, Z). Since measuring a qubit in *I* and *Z* bases yields the same result, LQ-reuse needs to prepare 4×3=12 variants for each sub-circuit. By contrast, if isomorphic sub-circuits are not reused, 24 sub-circuit variants need to be prepared. Besides, small-sized circuits have higher fidelity. Cutting a large-sized circuit into smaller isomorphic sub-circuits improves fidelity when evaluating a large-sized circuit on noisy QPUs. Reusing isomorphic sub-circuits can reduce the number of sub-circuits that need to be executed, but it may introduce noise biases in the final output distribution. In Section 5.5, we further discuss the impact of LQ-reuse on fidelity and show that LQ-reuse is practical.

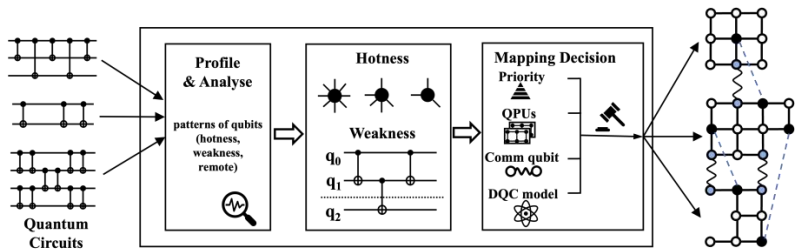

Figure 11. LarQucut's circuit mapping approach in a nutshell.

## 4 MAPPING SUB-CIRCUITS ACROSS QPUS

Through the cutting approach in Section 3, a large-sized quantum circuit is cut into several smaller sub-circuits, or a large circuit that has fewer interactions between qubit groups. This section introduces how to map these circuits across QPUs. Figure 11 shows the circuit mapping approach (LQ-mapping) in a nutshell. LQ-mapping first profiles the input circuits and identifies critical interaction patterns of the qubits, e.g., hotness and weakness. Then, it performs initial mapping using two mapping policies that work well for circuits with diverse patterns. Finally, it performs the mapping transition using a heuristic cost function that moves qubits with frequent interactions to the same DQC node. More details are as follows.

### 4.1 Profiling a Quantum Circuit

This component in LQ-mapping profiles and analyzes quantum circuits. First, LQ-mapping profiles all sub-circuits and identifies the "hotness" of each logical qubit. We have the term "hotness" to represent how frequently a specific qubit interacts with other qubits. The "hotness" of a specific qubit is the number of interactions it has with other qubits. Figure 12-(a) shows an example. The qubit $q_0$ has five interactive CNOT gates with other qubits; $q_1$ and $q_2$ have 3 and 2 interactive gates, respectively. We denote that $q_0$ is hotter than $q_1$, and $q_1$ is hotter than $q_2$. Leveraging hotness for qubits can be an effective way to reduce the number of remote operations in DQC, as we can avoid mapping the hot qubits and their neighbors across QPUs. Moreover, we find that in DQC, the "hotness" can change across different quantum chips. For example, suppose we have a large-sized circuit mapped across two quantum chips - nodes A and B. One hot qubit on node A interacts with a few qubits mapped on the same node. However, some other qubits interacting with this hot qubit are on node B. Therefore, the cost-effective way is to move/swap this hot qubit to node B using remote SWAP via the EPR pair, as shown in Figure 13. Second, we have another term, "weakness", which represents how frequently several groups of qubits interact with each other. For example, Figure 12-(b) illustrates that there are 3 CNOTs between $q_0$ and $q_1$, and only 1 CNOT between $q_1$ and $q_2$, indicating that the correlation between $q_1$ and $q_2$ is weaker. When two groups have weak interactions, there are fewer inter-node communications when the two groups are mapped onto two separate QPUs. The weakness between two groups of logical qubits is defined as 1 / (the number of inter-group two-qubit gates). A higher value of weakness indicates that the two groups are less correlated.

### 4.2 Initial Mapping Policies for Diverse Circuit Patterns

When a specific QPU does not have sufficient physical qubits to host all logical qubits, remote communications are necessary. LQ-mapping maps the circuit across QPUs in this case. Based on the hotness and weakness metrics, we have the hotness-mapping and weakness-mapping policies for mapping circuits with diverse patterns.

**Hotness-mapping.** Regarding the hotness metric, hotness-mapping first maps the hot logical qubits to physical qubits with high reliability, and then those qubits with weak interactions are mapped across QPUs. By doing so, the advantage is the reliable/robust physical qubits can be mapped with hot qubits. Quantum programs with hot logical qubits are prioritized in hotness-mapping. The policy finds a logical qubit with the highest hotness value and maps it onto a specific QPU. The physical qubit with the minimum value of 1 / (qubit degree × average CNOT reliability) is selected to map the logical qubit, where the average CNOT reliability is the average of (1 – CNOT error rate) for all CNOTs that can be executed using this qubit. The selected qubit has a high degree and high average reliability. Then, hotness-mapping maps the qubits that have interactions with this qubit. They are sorted in descending order of the number of interactions with the mapped qubit. For each of these qubits, hotness-mapping maps it onto the physical qubit that can maximize the total reliability of paths between this logical qubit and other mapped logical qubits. To calculate the reliability of each path, we multiply the reliability (i.e., 1 - CNOT error rate) of each CNOT on the path. Algorithm 1 shows this central logic. This process is repeated until all qubits are mapped.

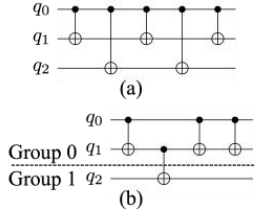
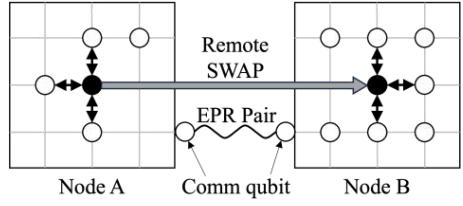

Figure 12. Hotness vs. weakness          Figure 13. Remote SWAP for hot qubit.

**Weakness-mapping.** Weakness-mapping first handles these qubits with weak interactions - maps them around communication qubits - and then maps these hot qubits. Via this approach, the advantage is ensuring the weak pairs can use communication qubits with a low swap overhead. As the number of communication qubits is limited, the policy always finds qubit partitions with a high value of weakness. For instance, for two qubit groups $G_1 = \{q_0, q_1\}$ and $G_2 = \{q_2, q_3\}$ with the highest value of weakness, $q_1$, $q_2$ are mapped to $QPU_1$ and $q_3$, $q_4$ are mapped to $QPU_2$. Therefore, the weakness-mapping policy can reduce remote gates by minimizing interactions across QPUs. During mapping, the policy always maps the qubits with a high value of weakness around the communication qubits on QPUs, reducing the SWAP overheads for using communication qubits. This process repeats until all QPUs' communication qubits are used. After handling the remote operations, the policy maps the remaining qubits onto QPUs. Weakness-mapping reduces the remote communication overheads if a quantum program has to have remote interactions. The main logic is in Algorithm 2.

---
Algorithm 1: Hotness-mapping
Input: Quantum programs; DQC typology; QPU calibration data
Output: Qubit mapping solution

---
BEGIN
1. Sort qubits in every program according to the hotness metric
2. Find the qubit A with a high hotness value
3. Map qubit A onto the robust physical qubit w/ robust neighbor links on a QPU //prefer to QPU w/ best-fit number of qubits
4. Map the qubits that interact with qubit A as the neighbors on this QPU
5. Map qubits belonging to one problem on a specific QPU
6. GOTO 1 and Repeat, and If the QPU does not have more qubit to satisfy the mapping, try to map the coming qubit onto another QPU
7. Always find a robust physical qubit w/ robust links to map
8. All qubits from all programs are mapped
9. Output the solution
END

---

---

Algorithm 2: Weakness-mapping
Input: Quantum programs; DQC typology; QPU calibration data
Output: Qubit mapping solution

---

BEGIN
1. Sort according to weakness metric for qubit groups, and find out the group A and B with the high value of weakness
2. Map the qubits pairs across the group A and B to communication qubits on different QPUs
3. Map the qubits in the group A that interact with the communication qubits to physical qubits close to it on one QPU
4. Map the qubits in the group B that interact with the communication qubits to physical qubits close to it on another QPU
5. Physical qubits on QPUs are mapped w/o the consideration of robust or not
6. GOTO Line 1 to repeat this process
7. All of the qubits are mapped, output the solution
END

---

### 4.3 Mapping Transition for Reducing EPR Pairs

During the mapping transition, LQ-mapping has a new heuristic approach to reduce remote communication overheads in DQC. It primarily handles the following cases. (1) For a specific logical qubit involved in consecutive remote CNOTs, LQ-mapping maps the qubit close to communication qubits, establishes an EPR pair, and executes the consecutive CNOTs by sharing the EPR pair. This case reduces remote communication overheads, as only one EPR pair is established. (2) For a logical qubit that interacts with many qubits on a remote node, LQ-mapping moves the logical qubit across nodes using remote SWAP to reduce remote communications. Additionally, even on a specific node, LQ-mapping's heuristic approach finds the shortest SWAP path that moves two logical qubits involved in one CNOT as a neighbor, making the CNOT executable.

In general, the heuristic approach handles the above cases by selecting the SWAP candidate with the minimum heuristic cost during the mapping transition. LQ-mapping takes a quantum circuit P as input. It builds the output hardware-compliant circuit by inserting SWAPs to P. In P, the front layer (F) denotes the set of quantum gates without unexecuted predecessors. When there are executable gates in F, LQ-mapping removes them from F and updates F. When all gates in F are not executable, LQ-mapping selects a SWAP candidate using the heuristic cost function. The SWAP candidates are SWAPs that involve at least one logical qubit used in F. Among them, the best SWAP that can handle the above cases is with the minimum heuristic cost. For case (1), the best SWAP has a low cost for remote communication. For case (2), the best SWAP has a low cost for moving logical qubits as neighbors. After the selected SWAP is conducted, the qubit mapping is changed, making some CNOTs in F executable. This process is repeated until all quantum gates are executable.

We have the above-mentioned heuristic function as below. When selecting a SWAP, LQ-mapping calculates a heuristic cost for each SWAP candidate. A lower heuristic cost indicates that the SWAP can reduce further SWAPs and remote communications. LQ-mapping employs NNC to calculate the heuristic cost. NNC is calculated as the sum of two parts for a quantum gate set. The first part is the minimum number of CNOTs required to move logical qubits of each CNOT as neighbors. For example, if one SWAP is required, the first part is 3, as one SWAP can be decomposed into three CNOTs [11,31,32]. For case 2 above, SWAPs that can move logical qubits closer are more easily selected, as they have a lower value in the first part. The second part is the remote communication overheads incurred by remote quantum gates, e.g., remote CNOTs and remote SWAPs. For instance, a remote gate using Cat-Comm incurs an additional overhead equivalent to 25 local CNOTs [9,10,11,30]. If consecutive remote gates can share one EPR pair in this gate set, the extra cost for all of them is 25. Otherwise, each remote gate has an additional cost of 25. For case 1 above, LQ-mapping shares one EPR pair among consecutive remote CNOTs, as doing so incurs lower additional costs in the second part. LQ-mapping calculates NNC for two quantum gate sets in the heuristic cost function: (1) CNOTs in the front layer (F), (2) an extended set (E) containing consecutive CNOTs using the same logical qubit. The heuristic cost function is -

$$H = \frac{1}{|F|}\text{NNC}(F) + 0.5\frac{1}{|E|}\text{NNC}(E).$$

The extended set E contains CNOTs from each logical qubit's first sequence of unexecuted consecutive CNOTs. For a logical qubit involved in consecutive CNOTs, mapping interactive qubits as its neighbors can reduce additional SWAPs. Therefore, LQ-mapping adds consecutive CNOTs to the extended set E to reduce SWAPs. The NNC values are normalized by the size of each quantum gate set. The NNC value for E has a weight parameter of 0.5, as CNOTs in F should be handled with a higher priority than those in E.

## 5 EVALUATIONS

### 5.1 Methodology

**Platforms.** We perform all experiments on a Linux 5.18.0 server equipped with due-socket Intel Xeon Gold 6338 CPUs and 256 GB RAM. Other software include Qiskit 1.0.2 [23] and Circuit Knitting Toolbox 0.7.0 [24].

**DQC models.** As there is no available DQC hardware open to the public, we use the noisy DQC systems simulated using Qiskit-aer [23] and Diskit [25] in evaluations. Specifically, we assume six DQC systems, as shown in Table 1. Each DQC system has twenty QPUs that have the same topology. These QPUs are arranged in the 1D nearest-neighbor topology in a DQC system, similar to the architecture used in the prior work [20]. For each QPU, two physical qubits are used as communication qubits. Remote EPR pairs can be established between neighboring QPUs using these communication qubits. The remaining physical qubits on the QPU are data qubits. Logical qubits can only be mapped on data qubits. For each QPU, we use real topology and noise information from IBM [23,34]. The noise is modeled using IBM Qiskit-aer [23].

**Metrics.** (1) Classical post-processing overheads. This metric represents the number of Kronecker product pairs [12] that need to be executed during the classical post-processing. It is calculated as $O(4^{k_1}6^{k_2})$, where k1 is the number of wire cuts, k2 is the number of gate cuts [12]. (2) Sampling overheads. The sampling overheads represent how many times the sub-circuits need to be executed [13]. It is calculated as $O(16^{k_1}9^{k_2})$. In quantum circuit cutting, the execution results of sub-circuits (i.e., samples) are combined to reconstruct the original circuit's result. Reusing isomorphic sub-circuits' execution results can reduce the sampling overheads. (3) The number of remote EPR pairs. Remote interactions between qubits mapped on different QPUs rely on remote EPR pairs. Each Cat-Comm requires one remote EPR pair. Consecutive remote gates that share the same logical qubit can be executed using one remote EPR pair. More remote EPR pairs increase errors and degrade the fidelity. (4) The absolute error between the expectation value and the ground truth of a quantum state measured on a Pauli basis. This metric represents the quality of the reconstructed results of the original circuit. The lower the absolute error, the higher the fidelity. The expectation value is the average value that would be obtained from measuring a quantum state on a Pauli basis many times. It is calculated as the average of all possible outcomes of a measurement weighted by their respective probabilities (i.e., 1 / number of samples). The absolute error is the difference between the expectation value of the experimental results and the ground truth.

**Competing approaches.** (1) CutQC [8]. CutQC is a circuit-cutting approach that cuts a quantum circuit into independent sub-circuits using wire cuts. It formulates the circuit-cutting problem as a mixed-integer programming problem and applies the Gurobi optimizer [19] to solve it. For CutQC, we set the maximal number of cuts to 500, the imbalance threshold to 500 gates, and allow up to 50 sub-circuits. (2) Circuit Knitting Toolbox (CKT) [24]. The circuit knitting toolbox is a quantum circuit-cutting approach that employs the best-first search. It uses wire cuts and gate cuts to cut the circuit into independent sub-circuits. (3) GP-Cat [36]. GP-Cat is a quantum circuit mapping approach in DQC environments. It employs Cat-Comm for inter-node communications, which is also used in LarQucut. GP-Cat maps the qubits using the static overall extreme exchange approach [37] to reduce inter-node communications.

Table 1. DQC systems used in evaluations.

| ID | QPUs | #Data qubits per QPU | #Comm. qubits per QPU |
|----|------|----------------------|-----------------------|
| 1 | 20×Manila | 3 | 2 |
| 2 | 20×Nairobi | 5 | 2 |
| 3 | 20×Melbourne | 13 | 2 |
| 4 | 20×Toronto | 25 | 2 |
| 5 | 20×Manhattan | 63 | 2 |
| 6 | 20×Washington | 125 | 2 |

**Benchmarks.** The benchmarks include Supremacy (SPM), Bernstein Vazirani (BV), Ripple-Carry Adder (RCA), Hardware-Efficient Ansatz (HWEA), and Quantum Fourier Transformation (QFT). Furthermore, to illustrate that LQ-reuse works well for circuits with repetitive interaction patterns, Greenberger-Horne-Zeilinger (GHZ) state and Linear-cluster (LC) state are used to evaluate LQ-reuse. Previous studies also use similar quantum circuits [8,9,17]. These circuits have diverse qubit interaction patterns and varying numbers of qubits. We evaluate them on DQC systems in Table 1. Using these circuits can show LarQucut's performance for large-sized circuits in complicated DQC environments.

## 5.2 Evaluations on LarQucut's Circuit-Cutting Effects

In this section, we compare LQ-hemicut with CutQC [8] and CKT [24] in terms of (1) the number of wire/gate cuts required to cut a large-sized quantum circuit, and (2) the classical post-processing overheads. To illustrate that LarQucut can reduce the number of cuts when it only uses critical cuts, we also evaluate cases where LarQucut always cuts a circuit into independent sub-circuits (denoted as LQ). Table 2 shows the experimental results. The DQC ID column in Table 2 shows the DQC systems used in evaluations. The #W and #G columns denote the number of wire cuts and gate cuts, respectively. The R% column denotes the percentage of remote gates removed by using LQ-hemicut. The classical post-processing overheads are shown in Figure 14.

Table 2. Evaluations on LarQucut's circuit-cutting effects.

| Workloads | #Qubits | QPUs | LQ-hemicut | | | LQ | | CutQC | CKT | |
|---|---|---|---|---|---|---|---|---|---|---|
| | | | #G | #W | R% | #G | #W | #W | #G | #W |
| SPM | 4 | Manila | 0 | 2 | 100% | 0 | 2 | 2 | 2 | 0 |
| BV | | | 0 | 1 | 100% | 0 | 1 | 1 | 1 | 0 |
| RCA | | | 1 | 0 | 100% | 1 | 0 | No solution | 1 | 0 |
| HWEA | | | 0 | 1 | 100% | 0 | 1 | 1 | 1 | 0 |
| QFT | | | 1 | 2 | 100% | 1 | 2 | No solution | 3 | 0 |
| SPM | 16 | Melbourne | 0 | 1 | 50% | 1 | 2 | 4 | 2 | 1 |
| BV | | | 0 | 1 | 100% | 0 | 1 | 1 | 0 | 1 |
| RCA | | | 0 | 2 | 100% | 0 | 2 | 2 | 0 | 2 |
| HWEA | | | 0 | 1 | 100% | 0 | 1 | 1 | 1 | 0 |
| QFT | | | 0 | 34 | 100% | 0 | 34 | 34 | 39 | 0 |
| SPM | 64 | Toronto | 0 | 13 | 94.2% | 0 | 15 | 15 | 37 | 0 |
| BV | | | 0 | 2 | 100% | 0 | 2 | 2 | 0 | 2 |
| RCA | | | 0 | 4 | 100% | 0 | 4 | 4 | 22 | 0 |
| HWEA | | | 0 | 2 | 100% | 0 | 2 | 2 | 2 | 0 |
| QFT | | | 0 | 507 | 100% | 0 | 507 | No solution | 1325 | 0 |
| SPM | 256 | Manhattan | 21 | 0 | 50% | 42 | 0 | 366 | 208 | 0 |
| BV | | | 0 | 4 | 100% | 0 | 4 | 4 | 193 | 0 |
| RCA | | | 22 | 0 | 50% | 44 | 0 | No solution | 44 | 0 |
| HWEA | | | 0 | 4 | 100% | 0 | 4 | 4 | 4 | 0 |
| SPM | 1024 | Washington | 75 | 0 | 49.7% | 149 | 0 | No solution | No solution | |
| BV | | | 0 | 8 | 100% | 0 | 8 | 406 | No solution | |
| RCA | | | 44 | 0 | 50% | 88 | 0 | No solution | 88 | 0 |
| HWEA | | | 0 | 8 | 100% | 0 | 8 | 406 | 341 | 0 |
| **Reduction of LQ-hemicut compared w/ other approaches** | | | | | | 12.2% | | 22.3% | 32.1% | |

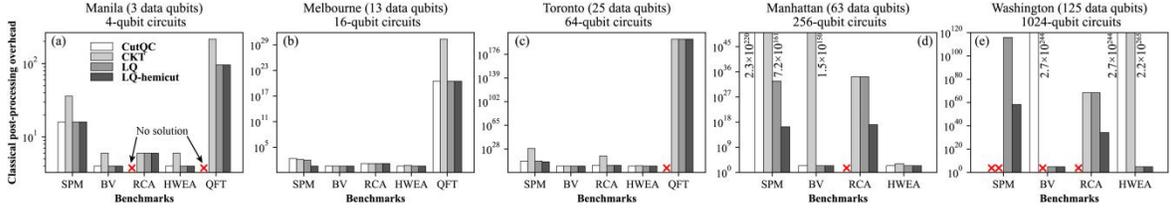

Figure 14. The classical post-processing overheads of CutQC, CKT, LQ, and LQ-hemicut when cutting circuits with 4, 16, 64, 256, and 1024 qubits. Red × marks indicate that no available solutions can be found. The QFT circuit incurs high overheads due to its dense pairwise qubit interactions. We only evaluate cases where QFT has 4, 16, and 64 qubits.

LQ-hemicut can effectively reduce the number of cuts required to cut a large-sized quantum circuit. As shown in Table 2, LQ-hemicut outperforms LQ, i.e., reducing the number of cuts by 12.2%, on average. The underlying reason is that LQ-hemicut does not have to cut a circuit into independent sub-circuits. It saves the cuts required to cut a large-sized circuit. For example, when cutting the 1024-qubit SPM circuit, LQ provides a cutting solution that uses 149 gate cuts. Yet, some cuts are unnecessary because they only remove a few inter-node communications. LQ-hemicut removes 74 non-critical gate cuts. Therefore, as shown in Figure 14-(e), LQ-hemicut reduces the classical post-processing overhead from $8.8\times10^{155}$ to $2.3\times10^{58}$. Moreover, LQ-hemicut can effectively reduce inter-node communications when circuits are mapped across QPUs. For circuits in Table 2, LQ-hemicut can remove 88.9% of remote gates, on average.

LQ-hemicut also outperforms CutQC and CKT, i.e., reducing the number of cuts by 22.3% and 32.1% compared with CutQC and CKT, respectively. Figure 14 shows that LQ-hemicut achieves the lowest classical post-processing overheads. Although LQ does not remove non-critical cuts and always cuts a circuit into independent sub-circuits, it can reduce the number of cuts by 18.2% and 23.4% compared with CutQC and CKT, on average, respectively. LarQucut (including LQ and LQ-hemicut) outperforms CutQC and CKT because (1) LarQucut can quickly identify critical cuts using the DQC topology information. By contrast, CutQC and CKT can not find critical cuts quickly and return a sub-optimal cutting solution that uses more cuts (e.g., CKT stops searching when it reaches the limited number of searching steps, which is 10000 by default). (2) LarQucut can find a better cutting solution by cooperatively using wire cuts and gate cuts. For example, when cutting the 16-qubit SPM circuit, LQ provides a cutting solution that uses one gate cut and two wire cuts, reducing the classical post-processing overheads by 62.5% and 33.3% compared with CutQC and CKT, respectively. Moreover, LarQucut can have solutions for more experiments. CutQC can not provide solutions for some circuits (e.g., 4-qubit QFT circuit) as it only uses wire cuts, which introduces new qubits when cutting a circuit. It can not provide a cutting solution where the qubit count in each sub-circuit does not exceed that of the QPU. CKT can not provide solutions for some large-sized quantum circuits, e.g., SPM and BV with 1024 qubits.

### 5.3 Evaluations on Execution Time

In this section, we evaluate the time required for sub-circuits execution and classical post-processing in DQC environments. We compare LarQucut with CKT because they employ the same approach in sub-circuits execution and classical post-processing. We show that LarQucut has lower time consumption due to its lower sampling overheads. CutQC employs the dynamic definition post-processing algorithm that can dynamically query solution states with high probabilities. Thus, it samples fewer sub-circuits and consumes less time than LarQucut and CKT. We evaluate 16-qubit circuits on Melbourne QPUs, and 32-qubit circuits on Toronto QPUs. Due to the exponential computational overheads of simulating quantum circuits on classical computers, the maximum size of the quantum circuits that can be evaluated is limited. Therefore, we only evaluate quantum circuits with up to 32 qubits. SPM is not used because it cannot return a valid circuit for 32 qubits. It can only generate circuits with a qubit count of $4^n$ ($n \geq 1$), based on the open-source implementation in the work [8]. The experimental results are in Table 3. The time limit for each circuit is set to 200 CPU hours. The "-" mark in the table indicates that the time required to run the circuit exceeds this time limit.

As shown in Table 3, LarQucut can reduce the time required for 16-qubit circuits by 36.1%, on average, compared with CKT. For 32-qubit circuits, LarQucut can have results for BV and GHZ within 200 CPU hours, but CKT can not have results for any circuits within the time limit. This is because LarQucut can reduce the sampling overheads by using fewer cuts. For example, CKT uses two wire cuts to cut the 16-qubit RCA circuit. It takes 758.9 seconds to execute the sub-circuits and 0.9 seconds in classical post-processing (i.e., 759.8 seconds in total). LarQucut uses one wire cut. It takes 671.8 seconds to execute the sub-circuits and 0.6 seconds in classical post-processing (i.e., 672.4 seconds in total).

Table 3. Evaluations on execution time.

| Benchmark | #Qubits | QPUs | CKT (seconds) | LarQucut (seconds) | Time reduction of LarQucut compared with CKT |
|---|---|---|---|---|---|
| BV | 16 | Melbourne | 71.7 | 53.5 | 25.4% |
| RCA | | | 759.8 | 672.4 | 11.5% |
| HWEA | | | 256.3 | 149.2 | 41.8% |
| GHZ | | | 63.3 | 20.7 | 67.2% |
| LC | | | 52.3 | 34.3 | 34.4% |
| QFT | | | - | - | - |
| BV | 32 | Toronto | - | 1083.5 | - |
| RCA | | | - | - | - |
| HWEA | | | - | - | - |
| GHZ | | | - | 5014.7 | - |
| LC | | | - | - | - |
| QFT | | | - | - | - |

### 5.4 Evaluations on LQ-reuse's Effects

In this section, we show that LQ-reuse can identify isomorphic sub-circuits in large-sized quantum circuits with repetitive qubit interaction patterns. We use workloads in Table 4, including (1) Quantum circuits that exhibit common repetitive qubit interaction patterns, i.e., GHZ, BV, and LC. We evaluate these circuits across 64-qubit, 256-qubit, 512-qubit, and 1024-qubit cases to show LQ-reuse's effects for large-sized quantum circuits. (2) Quantum circuits that exhibit fewer or no repetitive qubit interaction patterns, i.e., (SPM, RCA, and HWEA). Repetitive qubit interaction patterns can be found in SPM and RCA, but such patterns are less common. HWEA has several rotation blocks with arbitrary rotation angles. Therefore, no repetitive qubit interaction patterns can be found in HWEA. We evaluate these circuits across 64-qubit and 256-qubit cases. The DQC system used in evaluations has 20 Toronto QPUs (#4 in Table 1). The competing approaches include CutQC, CKT, and LQ (LarQucut that does not reuse isomorphic sub-circuits). Table 4 shows the number of wire/gate cuts required to cut the circuit. The #Isomorph column denotes the number of isomorphic sub-circuits identified by LC-reuse. The Time column denotes the time consumption for identifying isomorphic sub-circuits in the circuit.

Table 4. Evaluation on LQ-reuse's effects for circuits with repetative patterns.

| Type | Benchmark | #Qubits | LQ-reuse | | | | LQ | | CutQC | CKT | |
|---|---|---|---|---|---|---|---|---|---|---|---|
| | | | #G | #W | #Isomorph | Time (s) | #G | #W | #W | #G | #W |
| Circuits with repetitive qubit interaction patterns | GHZ | 64 | 0 | 2 | 0 | 0.2 | 0 | 2 | 2 | 2 | 0 |
| | BV | 64 | 0 | 2 | 0 | 2.5 | 0 | 2 | 2 | 0 | 2 |
| | LC | 64 | 0 | 2 | 0 | 12.8 | 0 | 2 | 2 | 2 | 0 |
| | GHZ | 256 | 0 | 11 | 10 | 0.9 | 0 | 10 | 97 | 10 | 0 |
| | BV | 256 | 0 | 11 | 10 | 6.1 | 0 | 10 | 97 | 231 | 0 |
| | LC | 256 | 0 | 12 | 12 | 11.2 | 17 | 0 | 18 | 10 | 0 |
| | GHZ | 512 | 0 | 21 | 20 | 3.1 | 0 | 20 | No solution | 20 | 0 |
| | BV | 512 | 0 | 21 | 20 | 16.6 | 0 | 20 | No solution | 487 | 0 |
| | LC | 512 | 0 | 25 | 25 | 34.8 | 2 | 18 | No solution | 21 | 0 |
| | GHZ | 1024 | 0 | 41 | 40 | 17.6 | 3 | 37 | No solution | 3 | 37 |
| | BV | 1024 | 0 | 41 | 40 | 77.8 | 0 | 40 | No solution | No solution | |
| | LC | 1024 | 0 | 51 | 51 | 43.1 | 10 | 30 | No solution | 42 | 0 |
| Circuits that exhibit fewer/none repetitive qubit interaction patterns | SPM | 64 | 14 | 0 | 2 | 0.4 | 14 | 0 | 16 | 37 | 0 |
| | RCA | 64 | 0 | 6 | 2 | 321.0 | 0 | 6 | 6 | 22 | 0 |
| | HWEA | 64 | 4 | 0 | 0 | 33.8 | 4 | 0 | 4 | 4 | 0 |
| | SPM | 256 | 95 | 0 | 2 | 164.9 | 95 | 0 | No solution | 193 | 0 |
| | RCA | 256 | 110 | 0 | 7 | 311.4 | 110 | 0 | No solution | 110 | 0 |
| | HWEA | 256 | 20 | 0 | 0 | 106.1 | 20 | 0 | No solution | 20 | 0 |

Table 5. Evaluations on LQ-reuse's impact on fidelity.

| #Qubits | QPUs | IDs | Workloads | Absolute error when n of all isomorphic sub-circuits are reused | |
|---|---|---|---|---|---|
| | | | | n = 0 | n = 1 |
| 8 | Manila | 1 | GHZ | 0.313 | 0.433 |
| | | 2 | BV | 0.299 | 0.392 |
| | | 3 | LCS | $2.0 \times 10^{-8}$ | $2.1 \times 10^{-7}$ |
| 16 | Melbourne | 4 | GHZ | 0.216 | 0.296 |
| | | 5 | BV | 0.327 | 0.375 |
| | | 6 | LCS | $5.0 \times 10^{-8}$ | $8.7 \times 10^{-7}$ |

For circuits that exhibit repetitive qubit interaction patterns, LQ-reuse can effectively identify and reuse isomorphic sub-circuits. For example, in Table 4, LQ-reuse can find 10, 20, and 41 isomorphic sub-circuits for BV circuits with 256, 512, and 1024 qubits, respectively. For circuits that exhibit fewer repetitive qubit interaction patterns (e.g., SPM and RCA), LQ-reuse can also find a few isomorphic sub-circuits. The execution results of isomorphic sub-circuits can be reused. For example, LQ-reuse cuts the 256-qubit BV circuit into 12 sub-circuits using 11 wire cuts. 10 out of these 12 sub-circuits are isomorphic. If only 8 of these isomorphic sub-circuits are evaluated and reused by the other 2 isomorphic sub-circuits. The sampling overheads would significantly reduce from $1.8 \times 10^{13}$ to $6.9 \times 10^{10}$. The number of isomorphic sub-circuits that reuse the execution results from others should be set carefully, as reusing isomorphic sub-circuits may introduce noise biases. In the next section, we will further evaluate LQ-reuse's impact on fidelity. Moreover, LQ-reuse works well for large-sized quantum circuits. The Time column in Table 4 shows the time required for LQ-reuse to identify isomorphic sub-circuits in the large-sized quantum circuit. LQ-reuse can identify isomorphic sub-circuits for circuits with up to 1024 qubits within minutes, which is acceptable in practice.

### 5.5 Evaluations on LQ-reuse's Impact on Fidelity

In this section, we evaluate the impact of LQ-reuse on fidelity. The experimental results are in Table 5. The benchmarks used in the evaluations are GHZ, BV, and LCS. We evaluate the 8-qubit circuits on Nairobi QPUs, and the 16-qubit circuits on Melbourne QPUs. For example, in the 16-qubit GHZ circuit on Nairobi QPUs (ID 4 in Table 5), LQ-reuse can find two isomorphic sub-circuits. When the execution results of isomorphic sub-circuits are not reused (i.e., n=0 in Table 5), the absolute error of GHZ is 0.216, and the sampling overhead is $16^3 = 4096$ (three wire cuts are used). When one of the isomorphic sub-circuits is reused (i.e., n = 1 in Table 5), the absolute error increases to 0.296, and the sampling overhead decreases to 256 (i.e., the sampling overhead is reduced by 93.75%). For other circuits, we can see similar phenomena. For example, for experiments with IDs 1, 2, and 5, reusing one isomorphic sub-circuit incurs an average of 35.5% higher absolute error but reduces the sampling overhead by 93.75%. Though LQ-reuse introduces noise biases, it is effective in reducing sampling overheads, making it possible to evaluate larger-sized quantum circuits, especially on today's quantum chips with a limited number of physical qubits.

### 5.6 Evaluations on LarQucut's Circuit-Mapping Effects

In this section, we compare LarQucut's circuit mapping approach (LQ-mapping) with GP-Cat [36] in terms of (1) the SWAPs required to map a circuit, (2) the number of EPR pairs consumed, and (3) the absolute error between the execu-tion results and the ground truth. We use the workloads in Table 6. The DQC ID column in Table 6 shows the DQC systems used for evaluating these circuits. The SWAPs, EPR, Depth, and Error columns in Table 6 denote the number of SWAPs, the number of consumed EPR pairs, the depth of the mapped circuit, and the absolute error, respectively.

Table 6. Evaluations on LarQucut's circuit-mapping effects.

| Workloads | #Qubits | QPUs | LQ-mapping | | | | GP-Cat | | | |
|---|---|---|---|---|---|---|---|---|---|---|
| | | | SWAPs | EPR | Depth | Error | SWAPs | EPR | Depth | Error |
| SPM | 4 | Manila | 2 | 2 | 27 | 0.016 | 2 | 2 | 30 | 0.022 |
| BV | | | 1 | 1 | 15 | 0.125 | 2 | 4 | 23 | 0.174 |
| RCA | | | 4 | 1 | 48 | 0.106 | 4 | 14 | 48 | 0.179 |
| HWEA | | | 0 | 1 | 10 | 0.020 | 0 | 2 | 10 | 0.147 |
| QFT | | | 3 | 3 | 35 | 4.9E-9 | 3 | 10 | 35 | 1.6E-7 |
| SPM | 16 | Melbourne | 15 | 10 | 68 | 0.002 | 14 | 13 | 51 | 0.037 |
| BV | | | 8 | 2 | 86 | 0.163 | 11 | 16 | 65 | 0.293 |
| RCA | | | 25 | 9 | 283 | 0.287 | 35 | 80 | 332 | 0.330 |
| HWEA | | | 0 | 2 | 90 | 0.030 | 10 | 13 | 40 | 0.053 |
| QFT | | | 69 | 18 | 268 | 4.9E-9 | 105 | 281 | 511 | 6.3E-7 |
| SPM | 64 | Toronto | 183 | 36 | 187 | - | 264 | 82 | 203 | - |
| BV | | | 108 | 13 | 371 | | 227 | 74 | 365 | |
| RCA | | | 241 | 68 | 1734 | | 244 | 70 | 1635 | |
| HWEA | | | 94 | 20 | 379 | | 194 | 62 | 354 | |
| QFT | | | 2446 | 468 | 2112 | | 2814 | 615 | 3345 | |
| SPM | 256 | Manhattan | 1821 | 359 | 684 | - | 2354 | 542 | 985 | - |
| BV | | | 1785 | 203 | 991 | | 1910 | 233 | 716 | |
| RCA | | | 1658 | 416 | 8720 | | 1934 | 542 | 8813 | |
| HWEA | | | 1180 | 268 | 2930 | | 1656 | 413 | 2754 | |
| QFT | | | 58103 | 8151 | 28245 | | 66616 | 13144 | 34633 | |

We can see that LQ-mapping uses fewer SWAPs and EPR pairs to map a circuit across QPUs. As shown in Table 6, LQ-mapping outperforms the GP-Cat, i.e., reducing the SWAPs, consumed EPR pairs, and circuit depth by 18.2%, 38.1%, and 6.3%, respectively. The underlying reasons include: (1) LQ-mapping identifies the qubit interaction patterns and uses two initial mapping policies adaptively to reduce inter-node communications. (2) The heuristic cost function has a lower cost value when consecutive remote gates can be executed sharing one EPR pair, which reduces the number of EPR pairs required. Besides, LQ-mapping achieves lower absolute error as it uses fewer EPR pairs. The Error column in Table 6 shows the absolute error of circuits with 4 and 16 qubits. Larger-sized circuits are not evaluated as they exceed the memory limit. Generally, LQ-mapping achieves a 39.4% reduction in terms of the absolute error compared with GP-Cat, on average.

### 5.7 Evaluations on Fidelity

In this section, we compare LarQucut with CutQC and CKT in terms of the absolute error (i.e., fidelity) when they are used to cut and map circuits in DQC environments. CutQC and CKT cut a quantum circuit and map the generated sub-circuits using the default mapping policy with the highest optimization level in Qiskit [23]. The workloads used in evaluations are in Figure 15. We use the DQC system that has 20 Manila QPUs (#1 in Table 1). LarQucut achieves the lowest absolute error on average. Compared with CutQC, LarQucut achieves 25.7%, 17.3%, and 11.3% lower absolute error for quantum circuits with 4, 8, and 16 qubits, on average, respectively. Compared with CKT, LarQucut achieves 27.7%, 22.1%, and 14.4% lower absolute error for the same circuits. These results show that LarQucut provides a more reliable result and incurs less fidelity reduction than other circuit-cutting approaches. The underlying reason is that LarQucut uses fewer cuts to cut a circuit. As errors accumulate during classical post-processing [17], higher classical post-processing overheads increase errors. LarQucut uses fewer cuts than other circuit-cutting approaches. Thus, the execution results of LarQucut are closer to the ground truth.

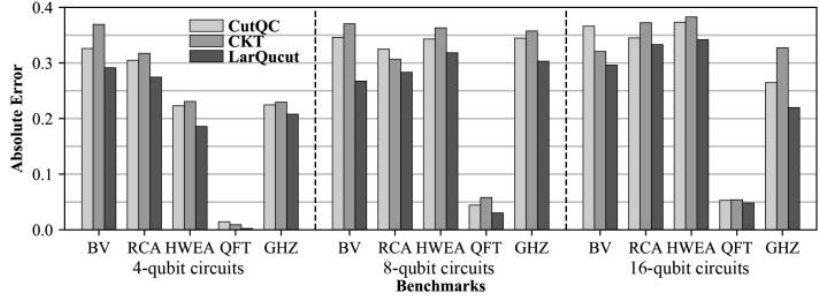

Figure 15. The absolute error of the expectation value compared with ground truth when cutting circuits with 4, 8, and 16 qubit circuits using CutQC, CKT, and LarQucut.

## 6   CONCLUSION

We present LarQucut, a new quantum circuit cutting and mapping approach for large-sized quantum circuits in DQC environments. It includes LQ-hemicut that can effectively remove inter-node communications without having to cut a circuit into independent sub-circuits; LQ-reuse that identifies and reuses isomorphic sub-circuits to reduce sampling overheads; and an adaptive quantum circuit mapping approach, which includes two initial mapping policies that work well for circuits with diverse patterns. Our experimental results show that LarQucut works well on large-sized quantum circuits in DQC environments. LarQucut cuts a large-sized quantum circuit into small-sized ones cost-effectively, making quantum computing with large-sized circuits feasible on DQC devices. LarQucut is a promising approach in the field of DQC, which can be essential in future quantum computing.

## ACKNOWLEDGMENT

We thank the reviewers, AE and EIC, for their valuable comments. Xinglei Dou, Zhuohao Wang and Pengyu Li are student members in Sys-Inventor Lab led by Lei Liu (PI). This project is supported by the National Key Research and Development Plan of China under Grant No. 2024YFB4504003, and the NSF of China under Grant No. 62072432. Lei Liu is the corresponding author.

## REFERENCES

[1]   Lov K. Grover, "A fast quantum mechanical algorithm for database search," in STOC, 1996.

[2]   Peter W. Shor, "Polynomial-Time Algorithms for Prime Factorization and Discrete Logarithms on a Quantum Computer," in SIAM Review, 1999.

[3]   Jacob Biamonte, Peter Wittek, Nicola Pancotti, Patrick Rebentrost, Nathan Wiebe, and Seth Lloyd, "Quantum machine learning," in Nature, 2017.

[4]   Yuval Boger, "Crossing the Quantum Threshold: The Path to 10,000 Qubits," https://www.hpcwire.com/2024/04/15/crossing-the-quantum-threshold-the-path-to-10000-qubits/

[5]   Chrissy Sexton, "Quantum computer built by Google can instantly execute a task that would normally take 47 years," https://www.earth.com/news/quantum-computer-can-instantly-execute-a-task-that-would-normally-take-47-years/

[6]   IBM, "IBM Unveils 400 Qubit-Plus Quantum Processor and Next-Generation IBM Quantum System Two," https://newsroom.ibm.com/2022-11-09-IBM-Unveils-400-Qubit-Plus-Quantum-Processor-and-Next-Generation-IBM-Quantum-System-Two

[7]   John Preskill, "Quantum Computing in the NISQ era and beyond," in Quantum, 2018.

[8]   Wei Tang, Teague Tomesh, Martin Suchara, Jeffrey Larson, and Margaret Martonosi, "CutQC: Using Small Quantum Computers for Large Quantum Circuit Evaluations," in ASPLOS, 2021.

[9] Anbang Wu, Hezi Zhang, Gushu Li, Alireza Shabani, Yuan Xie, and Yufei Ding, "AutoComm: A Framework for Enabling Efficient Communication in Distributed Quantum Programs," in MICRO, 2022.

[10] Anbang Wu, Yufei Ding, and Ang Li, "QuComm: Optimizing Collective Communication for Distributed Quantum Computing," in MICRO, 2023.

[11] Michael A Nielsen, and Isaac Chuang. "Quantum computation and quantum information", 2002.

[12] Aditya Pawar, Yingheng Li, Zewei Mo, Yanan Guo, Youtao Zhang, Xulong Tang, and Jun Yang, "Integrated Qubit Reuse and Circuit Cutting for Large Quantum Circuit Evaluation," in arXiv, 2312.10298.

[13] Sebastian Brandhofer, Ilia Polian, and Kevin Krsulich, "Optimal Partitioning of Quantum Circuits Using Gate Cuts and Wire Cuts," in Quantum Software, 2023.

[14] Tianyi Peng, Aram W. Harrow, Maris Ozols, and Xiaodi Wu, "Simulating Large Quantum Circuits on a Small Quantum Computer," in Physical Review Letters, 2020.

[15] Kosuke Mitarai, and Keisuke Fujii, "Constructing a virtual two-qubit gate by sampling single-qubit operations," in New Journal of Physical, 2021.

[16] Anbang Wu, Yufei Ding, and Ang Li, "CollComm: Enabling Efficient Collective Quantum Communication Based on EPR buffering," in arXiv, 2208.06724v2.

[17] Chong Ying, Bin Cheng, Youwei Zhao, He-Liang Huang, Yu-Ning Zhang, Ming Gong, Yulin Wu, Shiyu Wang, Futian Liang, Jin Lin, Yu Xu, Hui Deng, Hao Rong, Cheng-Zhi Peng, Man-Hong Yung, Xiaobo Zhu, and Jian-Wei Pan, "Experimental Simulation of Larger Quantum Circuits with Fewer Superconducting Qubits," in Physical Review Letters, 2023.

[18] Markus Reiher, Nathan Wiebe, Krysta M. Svore, Dave Wecker, and Matthias Troyer, "Elucidating reaction mechanisms on quantum computers," in Physical Sciences, 2016.

[19] "The Leader in Decision Intelligence Technology - Gurobi Optimization," https://www.gurobi.com/

[20] Davide Ferrari, Angela Sara Cacciapuoti, Michele Amoretti, and Marcello Caleffi, "Compiler Design for Distributed Quantum Computing," in Quantum Internet.

[21] Hezi Zhang, Hassan Shapourian, Keyi Yin, Alireza Shabani, Anbang Wu, and Yufei Ding, "MECH: Multi-Entry Communication Highway for Superconducting Quantum Chiplets," in ASPLOS, 2024.

[22] Turbasu Chatterjee, Arnav Das, Shah Ishmam Mohtashim, Amit Saha, and Amlan Chakrabarti, "Qurzon: A Prototype for a Divide and Conquer-Based Quantum Compiler for Distributed Quantum Systems," in SN Computer Science, 2022.

[23] Ali Javadi-Abhari, Matthew Treinish, Kevin Krsulich, Christopher J. Wood, Jake Lishman, Julien Gacon, Simon Martiel, Paul D. Nation, Lev S. Bishop, Andrew W. Cross, Blake R. Johnson, and Jay M. Gambetta, "Quantum computing with Qiskit," in arXiv, 2405.08810.

[24] Agata M. Branczyk, Almudena Carrera Vazquez, Daniel J. Egger, Bryce Fuller, Julien Gacon, James R. Garrison, Jennifer R. Glick, Caleb Johnson, Saasha Joshi, Edwin Pednault, C. D. Pemmaraju, Pedro Rivero, Seetharami Seelam, Ibrahim Shehzad, Dharmashankar Subramanian, Wei Tang, and Stefan Woerner, "Circuit Knitting Toolbox," https://github.com/Qiskit-Extensions/circuit-knitting-toolbox

[25] Anuranan Das, and Stephen DiAdamo, "diskit: Distributed QC for Qiskit," https://github.com/Interlin-q/diskit

[26] Davide Ferrari, Angela Sara Cacciapuoti, Michele Amoretti, and Marcello Caleffi, "Compiler Design for Distributed Quantum Computing," in Quantum Internet.

[27] Anabel Ovide, Santiago Rodrigo, Medina Bandic, Hans Van Someren, Sebastian Feld, Sergi Abadal, Eduard Alarcon, and Carmen G. Almudever, "Mapping quantum algorithms to multi-core quantum computing architectures," in ISCAS, 2023.

[28] Davide Ferrari, Stefano Carretta, and Michele Amoretti, "A Modular Quantum Compilation Framework for Distributed Quantum Computing," in IEEE Transactions on Quantum Engineering, 2023.

[29] Alpár Jüttner, and Péter Madarasi, "VF2++—An improved subgraph isomorphism algorithm," in Discrete Applied Mathematics, 2018.

[30] Marcello Caleffi, Michele Amoretti, Davide Ferrari, Daniele Cuomo, Jessica Illiano, Antonio Manzalini, and Angela Sara Cacciapuoti, "Distributed Quantum Computing: a Survey," in arXiv, 2212.10609.

[31] Gushu Li, Yufei Ding, and Yuan Xie, "Tackling the Qubit Mapping Problem for NISQ-Era Quantum Devices," in ASPLOS, 2019.

[32] Lei Liu, and Xinglei Dou, "QuCloud+: A Holistic Qubit Mapping Scheme for Single/Multi-programming on 2D/3D NISQ Quantum Computers," in ACM TACO, 2024.

[33] Junpyo Kim, Dongmoon Min, Jungmin Cho, Hyeonseong Jeong, Ilkwon Byun, Junhyuk Choi, Juwon Hong, and Jangwoo Kim, "A Fault-Tolerant Million Qubit-Scale Distributed Quantum Computer," in ASPLOS, 2024.

[34] IBM, "IBM Quantum Experience," https://quantum-computing.ibm.com/

[35] Ethan Bernstein, and Umesh Vazirani, "Quantum Complexity Theory," in SIAM Journal on Computing, 1997.

[36] Davide Ferrari, Angela Sara Cacciapuoti, Michele Amoretti, and Marcello Caleffi, "Compiler design for distributed quantum computing," in IEEE Transactions on Quantum Engineering, 2021.

[37] Jonathan M. Baker, Casey Duckering, Alexander Hoover, and Frederic T. Chong, "Time-sliced quantum circuit partitioning for modular architectures," in CF, 2020.

[38] Debasmita Bhoumik,∗ Ritajit Majumdar, Amit Saha, and Susmita Sur-Kolay, "Distributed Scheduling of Quantum Circuits with Noise and Time Optimization," in arXiv, 2309.06005.

[39] Joseph Clark, Travis S. Humble, and Himanshu Thapliyal, "GTQCP: Greedy Topology-Aware Quantum Circuit Partitioning," in QCE, 2023.

[40] Siwei Tan, Debin Xiang, Liqiang Lu, Junlin Lu, Qiuping Jiang, Mingshuai Chen, Jianwei Yin, "MorphQPV: Exploiting Isomorphism in Quantum Programs to Facilitate Confident Verification," in ASPLOS, 2024.